\documentclass[12pt]{article}
\usepackage{caption}
\usepackage{graphicx,latexsym,amssymb,amsmath}
\usepackage{tikz}
\usepackage{tabularx}
\usepackage{diagbox}
\usepackage{subfig}
\usepackage{mwe}
\definecolor{myblue}{RGB}{192,192,192}
\DeclareMathOperator*{\argmin}{arg\,min}
\usepackage[french,english]{babel}
\usepackage{enumerate}
\usepackage{here} 
\def\R{{\mathbb{R}}}

\def\P{{\mathbb{P}}}

\newcommand{{\convp}}{{\buildrel p\over\longrightarrow}}

\newcommand{{\Vs}}{{\cal V}}
\newcommand{{\Ps}}{{\cal P}}
\newcommand{{\Ss}}{{\cal S}}
\newcommand{{\Xs}}{{\cal X}}
\newcommand{{\Ls}}{{\cal L}}
\newcommand{{\Ns}}{{\cal N}}
\newcommand{{\Zs}}{{\cal Z}}
\newcommand{{\Fs}}{{\cal F}}

\newtheorem{Lemma}{Lemma}[section] 

\newtheorem{Proposition}{Proposition}[section]
\newtheorem{Theorem}{Theorem}[section]
\newcommand{\Poofend}{$\quad\Box{~}$}
\newenvironment{Proof}{\noindent {\em{\bf Proof:}}}{\Poofend\\}

\renewcommand{\theequation} {\arabic{section}.\arabic{equation}}
\renewcommand{\baselinestretch}{1.1}

\makeatletter
\renewcommand\theequation{\thesection.\arabic{equation}}
\@addtoreset{equation}{section}
\makeatother

\usepackage{natbib}

\begin{document}
\title{\bf A nonparametric instrumental approach to endogeneity in competing risks models}

\author{
{\large Jad B\textsc{eyhum}}
\footnote{ORSTAT, KU Leuven. Financial support from the European Research Council (2016-2021, Horizon 2020 / ERC grant agreement No.\ 694409) is gratefully acknowledged.}\\\texttt{\small jad.beyhum@kuleuven.be}
\and
\addtocounter{footnote}{2}
{\large Jean-Pierre F\textsc{lorens}}
\footnote{Toulouse School of Economics, Universit\'e Toulouse Capitole. Jean-Pierre Florens acknowledges funding from the French National Research Agency (ANR) under the Investments for the Future program (Investissements d'Avenir, grant ANR-17-EURE-0010).}
\\\texttt{\small jean-pierre.florens@tse-fr.eu}
\and
{\large Ingrid V\textsc{{an} K{eilegom}}$^*$}
\\\texttt{\small ingrid.vankeilegom@kuleuven.be}
}

\date{\today}

\maketitle

\begin{abstract}
This paper discusses endogenous treatment models with duration outcomes, competing risks and random right censoring. The endogeneity issue is solved using a discrete instrumental variable. We show that the competing risks model generates a nonparametric quantile instrumental regression problem. The cause-specific cumulative incidence, the cause-specific hazard and the subdistribution hazard can be recovered from the regression function. A distinguishing feature of the model is that censoring and competing risks prevent identification at some quantiles. We characterize the set of quantiles for which exact identification is possible and give partial identification results for other quantiles. We outline an estimation procedure and discuss its properties. The finite sample performance of the estimator is evaluated through simulations. We apply the proposed method to the Health Insurance Plan of Greater New York experiment.
\end{abstract}
\smallskip


\bigskip

\def\baselinestretch{1.3}

\newpage
\normalsize

\setcounter{footnote}{0}
\setcounter{equation}{0}
\section{Introduction}
Competing events are events that prevent the statistician from observing the time until the event of interest. A typical example is in biological studies with multiple causes of death. When the competing events are independent from the main outcome, then methods from the survival analysis literature that are designed for random right censoring can be used (Kaplan-Meier, Cox, ...). However, when the causes are dependent, a more careful analysis is necessary and only some features of the model can be identified. As a result competing risks problems can be seen as dependent censoring problems, which may arise if, for instance, some study participants decide not to attend a follow-up interview.

We consider a setting where the researcher is interested in the effect of a treatment on a randomly right-censored duration outcome in the presence of competing risks. The treatment is endogenous, it is not independent of the potential outcomes of the duration. In this case a naive analysis based on data conditional to treatment status could yield biased estimates, if, for example, treated study participants are the ones with the most positive treatment effect. Various approaches have been proposed in the literature to solve the endogeneity issue. Our method is based on an instrumental variable which is sufficiently dependent of the treatment but only affects the outcomes through the treatment. In a typical randomized experiment with binary treatment where there is noncompliance, a natural instrument is the treatment/control group assignment.

This paper focuses on the case where both the treatment and the instrument are categorical. We show that the competing risks model gives rise to a nonparametric nonadditive regression problem. The typical features of interest in competing risks models are functionals of the regression function, which we identify thanks to the instrumental variable. The present framework differs from the usual setting of the nonparametric nonseparable instrumental regression literature (\citet{CH, chernozhukov2006instrumental, C, wuthrich2020comparison}) because we allow for random right censoring (arising for instance from the end of the observation period) and competing risks (dependent censoring). The regression function cannot be identified for every value of the residual (which corresponds to quantiles of the distribution of the potential outcomes). The cause of identification failure can be censoring, competing risks or both. We single out the points at which the regression function is exactly identified and for the other quantiles we provide partial identification results. We discuss an estimation procedure and assess its performance through simulations. The strategy is applied to the Health Insurance Plan of Greater New York experiment.

When there are no competing risks, this problem has been thoroughly studied (\citet{AVdB, AVdB2, BR,chernozhukov2015quantile, frandsen2015treatment, tchetgen2015instrumental, li2015instrumental, chan2016reader, sant2016program, beyhum2021nonparametric}) with both parametric and nonparametric methods. In particular, \citet{beyhum2021nonparametric} also studies a nonparametric instrumental regression problem. The present paper makes two main contributions with respect to the latter works. First, we show how a competing risks model implies a nonparametric regression model. In the absence of competing risks, the duration model can naturally be formulated as a regression model, but in our case the link is far from obvious. Second, we tackle new identification and estimation challenges that arise uniquely because of the presence of competing risks. Note that the mechanism through which competing risks affect identification (non strict monotonicity of the regression function) are different from the effect of random right censoring (non identification of conditional survival functions).

Recent works have tackled the problem in the presence of competing risks. Unlike the present paper, \citet{kjaersgaard2016instrumental, zheng2017instrumental, ying2019two, martinussen2020instrumental} have semiparametric frameworks. 
 In the case where the treatment and the instrument are binary, \citet{richardson2017nonparametric, blanco2019bounds} study nonparametric models. Their approaches rely on a monotonicity assumption stating that there are no defiers (see \citet{angrist1996identification}). This is a restriction on the compliance of study participants to their treatment assignment, which has been criticized (see \citet{de2017tolerating}). Thanks to this condition, \citet{richardson2017nonparametric} identify the treatment effects on the cause-specific cumulative incidence functions for the population of compliers, that is agents whose treatment status follows treatment assignment. In \citet{ blanco2019bounds} bounds on the treatment effects for the population of compliers who experience the duration spell of interest are provided, they are valid when the censoring is dependent, which could correspond to a competing risk. In contrast, the present paper makes a rank invariance assumption. It restricts the distribution of potential outcomes, such that ranks (in a sense that is given in Appendix A) between subjects cannot be reversed by the treatment. This condition allows us to identify and estimate the treatment effects on the full population, which is arguably more interesting when one weighs whether or not to make a treatment compulsory. See \citet{wuthrich2020comparison} for a discussion of the trade-off between the two strategies.

The paper is organized as follows. In Section \ref{sec.main_model}, we outline the model. Then, identification is discussed in Section \ref{sec.ident}. Estimation is studied in Section \ref{sec.est}. In Section \ref{sec.sim}, we present simulations. The method is applied to the Health Insurance Plan of Greater New York experiment in Section \ref{sec.app}. Finally, concluding remarks are given in Section \ref{sec.conc}.

\section{The model}\label{sec.main_model}

\subsection{Structural model}
\label{sec:model}

The objective of this paper is to analyze treatment models when the outcome of the treatment is a duration time that is subject to competing risks and random right censoring. Suppose that there are two latent durations $T_j,j=1,2$. For $j=1,2$, let $U_j$ be a residual which represents the dependence of $T_j$ on unobserved heterogeneity. Without loss of generality, the distribution of $U_j$ is normalized to be unit exponential. The dependence of $U_1$ and $U_2$ is left unrestricted (hence, the independent case is nested by our model). We assume that there exists a continuous and strictly increasing mapping $\psi_j: \R_+\mapsto \R_+$ such that
$$T_j=\psi_j(U_j);\ j=1,2.$$
The objects of interest fail at time $T=\min(T_1,T_2)$ and the cause of this failure is $E=\min\{\argmin_{j=1,2}T_j\}$. Let us now introduce the probability of failure from cause $1$, that is $p=\P(E=1)$, and the conditional cumulative distribution function of $U_j$ given $E=j$, that is $F_j(u_j)=\P(U_j\le u_j|E=j)$ for $u_j\in \R_+$. We assume that $F_j,j=1,2$ is continuous and strictly increasing on the support of $U_j$ given $E=j$. We define the residual
\begin{equation}\label{defU} U= pF_1(U_1)I(E=1) + (p+(1-p)F_{2}(U_2))I(E=2),\end{equation}
where $I(\cdot)$ is the indicator function. It can be shown that the distribution of $U$ is uniform on $[0,1]$ (See Lemma A.1 in Appendix A.1). The variable $T$ is generated by $U$ through the following relationship:
\begin{equation}\label{modelwoz}T=\left\{\begin{array}{cc} \psi_1\Big((F_{1})^{-1}\Big(\frac{U}{p}\Big)\Big) &\text{ if } U< p\\
 \psi_2\Big((F_{2})^{-1}\Big(\frac{U-p}{1-p}\Big)\Big) &\text{ if $U>p$.}\end{array}\right.\end{equation}
Remark that the model does not say anything about what happens at $U=p$. On this event, it is possible that either $E=1$ and $E=2$. Because $\P(U=p)=0$, this is innocuous and ensures the symmetry of the model.

Let us now introduce a categorical treatment $Z$ with support $\{z_1,\dots,z_L\}$. We are interested in the potential outcomes (see \citet{rubin2005causal}) of $(T,E)$ under treatment status $z\in\{z_1,\dots,z_L\}$, which are denoted by $(T(z),E(z))$. We assume that the potential outcomes follow a model of the type \eqref{modelwoz}, in the sense that for $z\in\{z_1, \dots,z_L\}$ there exists $p_z\in(0,1)$  and continuous and strictly increasing mappings $\varphi_1(z,\cdot):[0,p_z)\mapsto \R_+$ and $\varphi_2(z,\cdot) :(p_z,1]\mapsto \R_+$ such that
\begin{equation} \label{model}T(z)=\left\{\begin{array}{cc} \varphi_1(z,U)&\text{ if } U<p_z\\
 \varphi_2(z,U) &\text{ if } U>p_z\end{array}\right.;\ E(z)=\left\{\begin{array}{cc} 1&\text{ if } U<p_z\\
 2 &\text{ if } U>p_z\end{array}\right.,
\end{equation}
and $T=T(Z),E=E(Z)$. When there are no $Z$, \eqref{model} becomes \eqref{modelwoz}.
Remark that $p_z=\P(E(z)=1)$ since $U$ is uniform on $[0,1]$, and that the residual $U$ does not vary among treatment statuses, this is the usual rank invariance assumption from the Econometrics literature (see \citet{dong2018testing}). In Appendix A.2 we argue that this assumption allows for a wide variety of treatment effects.

This paper is concerned with identification and estimation of some features of the model (see Section \ref{quantities}) when $Z$ is endogenous and $T=T(Z)$ is randomly right censored. We treat the confounding issue thanks to an instrumental variable (henceforth, IV). 
Formally, $Z$ and $U$ are dependent but we possess a categorical instrumental variable $W$ with support $\{w_1,\dots,w_K\}$ such that $U$ and $W$ are independent. There also exists a censoring variable $C$ with support in $\R_+$ such that we observe $Y=\min(T,C)$ and the censoring indicator $\delta=I(T\le C)$. 

Our model can be regarded as a quantile regression model. Indeed, if we introduce the random variable
$$T^j(z)=T(z)I(E(z)=j)+\infty I(E(z)\ne j)$$
for $z\in\{z_1,\dots,z_L\}$, then we have $T^j(z)=\varphi^j(z,U)=\varphi^j_z(U),$ where $\varphi^j_z(\cdot):[0,1]\mapsto [0,\infty]$ is such that, for all $u\in[0,1], u\ne p_z$,
$$\varphi^1_z(u)=\left\{\begin{array}{cc} \varphi_1(z,u)&\text{ if } u< p_z\\
 \infty &\text{ if } u> p_z\end{array}\right.;\ \varphi^2_z(u)=\left\{\begin{array}{cc}\infty&\text{ if } u< p_z\\
 \varphi_2(z,u) &\text{ if } u> p_z.\end{array}\right.$$
The quantity $\varphi_z^1(u)$ is the $u$-quantile of the distribution of $T^1(z)$ and, hence, the model $T^1=\varphi^1(Z,U)$ can be seen as a IV model of quantile treatment effects as in \citet{CH}. There are however two distinguishing differences with the usual model of \citet{CH}.  
First, the variable $T$ is randomly right censored. Second, the mapping $\varphi^1_z(\cdot)$ is not strictly increasing on all of its support. These differences pose additional identification and estimation issues which we tackle in this paper. 

Throughout the paper, we suppose that the distribution of $U$ given $\{Z=z,W=w\}$ is continuous, with strictly positive density on $[0,u_{z,w})$, where 
$$u_{z,w}=\sup\{u\in\R_+:\ \P(U\le u|Z=z,W=w)<1\}$$
is the upper bound of the support of the distribution of $U$ given $\{Z=z,W=w\}$.
Remark that this allows the support of the latter distribution to differ from $[0,1]$. We conclude the presentation of the model $T^1=\varphi^1(Z,U)$ by summarizing the underlying assumptions.
\begin{itemize}
\item[\textbf{(M)}]\begin{itemize}
\item[(i)] $\varphi^1_z$ is continuous and strictly increasing and $[0,p_z)$ and equal to $\infty$ on $(p_z,1]$;
\item[(ii)] $U$ has a uniform distribution on the interval $[0,1]$ and independent of $W$;
\item[(iii)] The distribution of $U$ given $\{Z=z,W=w\}$ is continuous, with strictly positive density on $[0,u_{z,w})$.
\end{itemize}
\end{itemize}
Note that no assumptions on $\varphi^2$ are needed to identifiy $\varphi^1$.

\subsection{Quantities of interest} \label{quantities}

Let us now discuss the features that we seek to recover. It is well-known that it is not possible to identify the distribution of the latent durations in a competing risks model (see \citet{tsiatis1975nonidentifiability}). However, some characteristics which we introduce below are identified when $Z$ is exogenous (see \citet{geskus2020competing}). For $z\in\{z_1,\dots,z_L\}$, we introduce the left continuous inverse of the function $\varphi^j_z$, that is $(\varphi^j_z)^{-1}:\ t\in[0,\infty]\mapsto \inf\{u\in [0,1]:\varphi^j_z(u)\ge t\}$. Let us define the cause-specific cumulative incidence function at time $t\in\R_+$ for cause $j$ under treatment status $z$, \begin{equation}\label{cumh}F_{z}^j(t)=\P(T^j(z)\le t)=(\varphi^j_z)^{-1}(t).\end{equation}
This equation implies that for $u\in[0,1]$, $\varphi^j_z(u)$ is the $u$-quantile of the subdistribution function $F^j$. Remark that $F_{z}^j(t)$ is a structural function and therefore it is different from the conditional probability $\P(T^j\le t|Z=z)=\P(T\le t, E=j|Z=z)$.
We also introduce the subdistribution hazard at time $t\in\R_+$ for cause $j$ under treatment status $z$, that is
\begin{equation}\label{subh}h_{z}^j(t)=\lim\limits_{dt \to 0}\frac{\P(t\le T^j(z) \le t+dt)}{dt \, \P(T^j(z)  \ge t)}=\frac{((\varphi_z^j)^{-1})'(t)}{1-(\varphi_z^j)^{-1}(t)}.\end{equation}
It is the hazard rate of the subdistribution function $F_{z}^j(t)$, i.e. $h_{z}^j(t)=f_{z}^j(t)/(1-F_{z}^j(t))$, where $f_{z}^j(t)$ is the derivative of $F_{z}^j(t)$ at $t$. Let also the cause-specific hazard at time $t$ for cause $j$  under treatment status $z$ be
\begin{equation}\label{cauh}\lambda_z^j(t)=\lim\limits_{dt \to 0}\frac{\P(t\le T^j(z) \le t+dt)}{dt \, \P(T(z)\ge t)}=\frac{((\varphi_z^j)^{-1})'(t)}{1-(\varphi^1_z)^{-1}(t)-(\varphi^2_z)^{-1}(t)}.\end{equation}
We are interested in the effect of the treatment status $z$ on these three quantities, that is the treatment effects. Consider, for instance, the $u$-quantile treatment effect on the subdistribution of the duration until failure from cause $1$, that is $\varphi_1^1(u)-\varphi_0^1(u)$. As is clear from equations \eqref{cumh}, \eqref{subh}, and \eqref{cauh}, these various treatment effects are identified from knowledge of $\varphi^j, j=1,2$. Therefore, without loss of generality, we focus on identification and estimation of $\varphi^1$ in the remainder of the paper.

Remark that the fact that we assumed that there are only two causes is not restrictive. Indeed, if there are $J\ge 3$ event types, the treatment effects for cause $j^*\in\{1,\dots,J\}$ can be identified from $\varphi^1$ if we label an exit from cause $j^*$ as $j=1$ and an exit from any other cause as $j=2$. Applying this modeling for each cause, that is $J$ times, allows to recover the treatment effects for all causes.

\section{Identification}
\label{sec.ident}
\subsection{System of equations}\label{sy}
In this section, we discuss identification of $\varphi^1$ when the distribution of the observables $(Y,\delta E, Z,W)$ is known. As usual in nonparametric instrumental regression, the model generates a nonlinear system of equations. Let $p^1=\min_{\ell=1}^Lp_{z_\ell}$. For $u\in[0,p^1)$, $(\varphi^1_{z_\ell}(u))_{\ell=1}^L$ is a solution to the following system of equations in $\theta\in \R^L$:
\begin{equation}\label{Sy}
\sum_{\ell=1}^LS^1(\theta_\ell,z_\ell|w_k)=1- u\, \text{ for } k=1,\dots,K,
\end{equation}
where $S^1(t,z|w) = \P(T^1 \ge t,Z=z|W=w)$. This property holds because
\begin{eqnarray}
\notag\sum_{\ell=1}^L S^1(\varphi_{z_\ell}^1(u),z_\ell|w_k)&=& \sum_{\ell=1}^L \P(T^1\ge \varphi_{z_\ell}^1(u),Z=z_\ell|W=w_k)\\
\notag &=& \sum_{\ell=1}^L \P(\varphi_{z_\ell}^1(U)\ge \varphi^1_{z_\ell}(u),Z=z_\ell|W=w_k)\\
\label{change} &=& \sum_{\ell=1}^L \P(U\ge u ,Z=z_\ell|W=w_k) \\
\notag&=& \P(U\ge u|W=w_k) = 1-u,
\end{eqnarray}
where \eqref{change} holds because $\varphi_z^1(\cdot)$ is strictly increasing on $[0,p^1)$. 
Usually, one would simply make assumptions ensuring that this system has a unique solution. However, the particular context of this paper brings two additional difficulties. First, $S^1(\cdot,z|w)$ may not be identified on all of its support because of censoring. Second, \eqref{Sy} is only valid for $u\in[0,p^1)$ because $\varphi^1_z(\cdot),z=z_1,\dots,z_L$ are not strictly increasing on all $[0,1]$, which is a direct consequence of the presence of competing risks. These two difficulties restrict the set of values of $u$ for which it is possible to identify $(\varphi_{z_\ell}(u))_{\ell=1}^L$ thanks to \eqref{Sy}. In the next two subsections, we characterize the latter set.

\subsection{The role of censoring}

We work throughout the paper with the usual independent censoring assumption:
\begin{itemize}
\item[\textbf{(C1)}] $C$ is independent of $(T,E)$ given $Z,W$.
\end{itemize}
Let us also define the upper bounds of the support of $C$ conditional on $\{Z=z,W=z\}$ and $\{Z=z\}$, that is
\begin{align*}
c_{z,w}&=\sup\{t\in\R_+:\ \P(C\le t|Z=z,W=w)<1\};\\
c_{z}&=\sup\{t\in\R_+:\ \P(C\le t|Z=z)<1\}.
\end{align*}
We make the assumption that the upper bound of the censoring time only depends on the treatment status:
\begin{itemize}
\item[\textbf{(C2)}] For all $z\in\{z_1,\dots,z_L\}$ and $w\in\{w_1,\dots,w_K\}$, we have $c_{z}=c_{z,w}$.
\end{itemize}
This hypothesis simplifies the analysis but could be relaxed. It is likely to hold when the follow-up scheme of the study only depends on the treatment status. Moreover, in this paper we assume that $T$ depends on $W$ only through $Z$, therefore it makes sense to make the same assumption about $C$.
Let $$S^1(t|z,w)=\P(T^1\ge t|Z=z, W=w)=\P(T\ge t, E=1|Z=z, W=w)$$
be the cause-1-specific survival function conditional to $\{Z=z,W=z\}$.  We have $S^1(t|z,w)=1-\int_0^t S(s|z,w)\lambda^1(s|z,w) ds$, where
\begin{align*}S(s|z,w)&= \P(T\ge s|Z=z,W=w)\\
\lambda^1(s|z,w) &=\lim_{ds\to 0} \frac{\P(T\in[s,s+ds], E=1|T\ge s, Z= z,W=w)}{ds}
\end{align*}
are the survival function of $T$ conditional on $Z=z, W=w$ and the cause-1-specific hazard rate given $Z=z, W=w$, respectively. The function $S(\cdot|z,w)$ is identified on $[0,c_z]$ by standard arguments from the survival analysis literature.
Moreover, remark that,  for all $t$ in $[0,c_{z})$, we have 
\begin{align*}&\lim_{dt\to 0}\frac{\P(Y\in[t,t+dt], \delta E=1|Y\ge t,Z=z, W=w)}{dt}\\
 &=\lim_{dt\to 0}\frac{\P(Y\in[t,t+dt], \delta=1, E=1, C\ge t, T\ge t|Z= z,W=w)}{dt \, \P(C\ge t,T\ge t| Z=z,W=w)}\\
 &=\lim_{dt\to 0}\frac{\P(T\in[t,t+dt], C\ge t, E=1|Z= z,W=w)}{dt \, \P(C\ge t,T\ge t |Z=z,W=w)}\\
 &=\lim_{dt\to 0}\frac{\P(T\in[t,t+dt], E=1|Z= z,W=w)\P(C\ge t|Z=z,W=w)}{dt \, \P(C\ge t |Z=z,W=w)\P(T\ge t |Z=z,W=w)}=\lambda^1(s|z,w) .
\end{align*}
Hence, $\lambda^1(s|z,w)$ is identified on $[0,c_z)$ and so are $S^1(\cdot|z,w)$ and $S^1(t,z|w)=S^1(\cdot|z,w)\P(Z=z|W=w)$. Next, we have two cases. In order to distinguish them, we introduce the (finite or infinite) upper bound of the support of the distribution of $T$ given $\{E=1, Z=z,W=w\}$:
$$t^1_{z,w}=\sup\{t\in\R_+:\ \P(T\le t|E=1,Z=z,W=w)<1\} .$$
If $c_z\le t^1_{z,w}$, then $S^1(\cdot,z|w)$ is not identified outside $[0,c_z]$. If instead  $c_z>t^1_{z,w}$, then the event $\{Y> t^1_{z,w},  \delta E\le 1, Z=z, W=w\}$ would have measure $0$ which would imply $\P\left(T> t^1_{z,w}, E=1, Z=z, W=w\right)=0$  leading to $S^1(t^1_{z,w},z|w)=S^1(\infty, z|w)$ and, therefore, $S^1(\cdot,z|w)$ is identified on $[0,\infty]$.
Let 
$$u_C=\sup\{u\in [0,1]:\ \varphi_{z_\ell}^1(u)< c_{z_\ell}\text{ or  $\varphi_{z_\ell}^1(u)=\infty$ for all $\ell=1,\dots,L$}\}.$$ 
The above analysis implies that for $u> u_C$, we do not know the mapping on the left hand side of \eqref{Sy} at the point $\theta= (\varphi_{z_\ell}(u))_{\ell=1}^L$. As a result, \eqref{Sy} cannot point identify $ (\varphi_{z_\ell}(u))_{\ell=1}^L$ for $u> u_C$.

\subsection{The role of the selection effect}\label{rse}

The set for which \eqref{Sy} has the potential to deliver point identification can be further reduced. Let us introduce
$$t^1_z=\sup\{t\in\R_+:\ \P(T\le t|E=1,Z=z)<1\}.$$
The quantity $t^1_z$ is the maximum of $t^1_{z,w}$ over $w\in\{w_1,\dots,w_K\}$. The mapping $S^1(\cdot, z|w)$ is flat on $[t^1_{z,w},\infty]$ by definition of $t^1_{z,w}$. Hence, for all $w\in\{w_1,\dots,w_k\}$, $S^1(\cdot, z|w)$ is constant on $[t^1_{z},\infty]$. Therefore, if $\varphi_z^1(u)$ belongs to $[t^1_{z},\infty]$, then $(\varphi^1_z(u))_{\ell=1}^L$ cannot be identified by the system \eqref{Sy}.  Because it is sufficient that this happens for one value of $z$ for identification to break down, it is not possible to identify $(\varphi^1_z(u))_{\ell=1}^L$  with \eqref{Sy} for all $u\ge u_E$, where $u_E=\min_{\ell=1}^L(\varphi_{z_\ell}^1)^{-1}(t^1_{z_\ell})$. The definition of $t^1_z$ ensures that $\varphi^1_z$ is invertible on $[0,t^1_z)$ and, hence, that $u_E$ is properly defined. 

It turns out that $u_E\le p^1$. Indeed, the support of the distribution of $T$ given $\{E=1,Z=z\}$ is equal to the support of the distribution of $\varphi_z^1(U)$ given $\{E=1,Z=z\}$. Since on the event $\{E=1,Z=z\}$ we have $U\le p_z$, we obtain that the support of the distribution of $T$ given $\{E=1,Z=z\}$ is included in the image of $\varphi_z^1(\cdot)$ on $[0,p_z]$. Hence, we have $t^1_z\le \varphi^1_z(p_z-)$,  where for a mapping $f:\R_+\to \R_+$ and $t\in\R_+$, $f(t-)$ is the left limit of $f$ at $t$. Therefore $(\varphi^1_z)^{-1}(t^1_z)\le p_z$ because $\varphi^1_z$ is strictly increasing on $[0,p_z]$. This leads to $u_E \le p^1$ as claimed. 

Notice that if $t^1_z$ were to be equal to $\varphi^1_z(p_z-)$, we would have that $(\varphi_z^1)^{-1}(t^1_z)=p_z$ and, hence, $u_E=p^1$. The latter happens when the upper bound of the support of $U$ given $Z=z$ is $p_z$. This shows that the fact that the system of equations cannot identify  $(\varphi^1_{z_\ell}(u))_{\ell=1}^L$ for $u\in(u_E,p^1)$ is due to the dependence of $U$ and $Z$, that is the selection into treatment.

At the moment, we have that the system \eqref{Sy} can only identify $(\varphi^1_{z_\ell}(u))_{\ell=1}^L$ for $u\in [0,u_Y)$ with $u_Y=   u_E\wedge u_C$ (or $u\in[0,u_Y]$ if $u_C<u_E$). Remark that even if the system \eqref{Sy} may not have a unique solution at $u_Y$, $(\varphi^1_{z_\ell}(u_Y))_{\ell=1}^L$ can always be identified by continuity as long as identification holds on $[0,u_Y)$. By definition $(\varphi^1_{z_\ell}(u))_{\ell=1}^L\in  \prod_{\ell^=1}^L[0,y^1_{z_\ell})$ for these values of $u$, where $y^1_z=t^1_{z_\ell}\wedge c_{z_\ell}$.

\subsection{Point identification}\label{pi}

Now, we make a strong conditional completeness assumption similar to that in Appendix C of \citet{CH} which ensures that \eqref{Sy} has a single solution $(\varphi^1_{z_\ell}(u))_{\ell=1}^L\in  \prod_{\ell^=1}^L[0,y^1_{z_\ell})$. We follow the presentation of Appendix A in \citet{FFV} of this condition. Assume that the density of $(U,Z)$ given $W$ is perturbed in the direction of a function $\Delta \in\mathcal{P}$, where $\mathcal{P}$ is the set of mappings $\Delta:\{z_1,\dots, z_L\}\times\R_+\mapsto \R_+$  such that $\varphi_z^1(u)+\Delta_z(u)\in[0,y^1_z)$ and $(\varphi_z^1)'(u)+\Delta_z'(u)>0$ for all $u\in[0,u_Y)$. Let $f^1(t,z|w)=-\frac{\partial S^1}{\partial t}(t,z|w)$. For $\mu \in[0,1]$, let us define $$
g_{\mu,\Delta}(u,z|w)=((\varphi_z^1)'(u)+\mu(\Delta_z)'(u))f^1(\varphi_z^1(u)+\mu\Delta_z(u),z|w).
$$
The main identification assumption is
\begin{itemize}
\item[\textbf{(G)}] If $\rho_\mu:\{z_1,\dots, z_L\}\times\R_+\mapsto \R $ is such that $$\int_0^1\sum_{\ell=1}^L \rho_\mu(z_\ell,u)g_{\mu,\Delta}(u,z_\ell|w)  d\mu=0$$ for all $u\in[0,u_Y)$, $w\in\{w_{1},\dots,w_K\}$ and $\Delta\in\mathcal{P}$, then $\rho_\mu \equiv 0$ for all $\mu \in[0,1]$.
\end{itemize}
It can be interpreted as a  strong conditional completeness condition of $Z$ and a uniform random variable $\mu$, given $(W,U)$ for the distribution $g$. In the case where $Z$ and $W$ are binary with support $\{0,1\}$ a simpler identification condition can be given. By Theorem 2 in \citet{CH}, it suffices to assume
\begin{itemize}
\item[\textbf{(G')}]  The density of $U$ given $(Z,W)$ is continuous and the matrix 
$$G(t)=\left(\begin{array}{cc}f^1(t_1,0|0)&f^1(t_2,1|0) \\
f^1(t_1,0|1)&f^1(t_2,1|1)
\end{array}\right)$$
has rank $2$ for all $t=(t_1,t_2)^\top\in[0,y^1_0)\times [0,y^1_1)$.
\end{itemize}
Assuming that $f^1(\cdot,z|w)$ is continuous, this full rank condition implies that the determinant of $G(t)$ is either $>0$ or $<0$ for all $t\in[0,y^1_0)\times [0,y^1_1)$, that is
$$\frac{f^1(t_2,1|1)}{f^1(t_1,0|1)}> \frac{f^1(t_2,1|0)}{f^1(t_1,0|0)}$$
 or the same inequality with $<$ instead of $>$. Following the semantic of \citet{CH}, this is a monotone likelihood ratio condition: the instrument increases (or decreases) the probability of being treated ($Z=1$) for all levels of outcomes $t\in[0,y^1_0)\times [0,y^1_1)$. In the case of one-sided noncompliance, where $\P(Z=1|W=0)=0$, the condition is trivially satisfied as long as $\P(Z=1|W=1)>0$.

To conclude on point identification, we need to be sure that $u_Y$ is identified because otherwise we are not able to know for which value of $u$ the quantity $(\varphi^1_{z_\ell}(u))_{\ell=1}^L$ is identified. Remark that $y^1_z=t^1_{z}\wedge c_{z}$ is identified because it is the upper bound of the support of the distribution of $Y$ given $\{E=1, Z=z\}$. Hence, $u_Y=u_E\wedge u_C=\min_{\ell=1}^L(\varphi_{z_\ell}^1)^{-1}(t^1_{z_\ell}\wedge  c_{z_\ell}-)$ is identified (Under Assumption (G), one can solve the system \eqref{Sy} for increasing values of $u$ until the left limit of $\varphi^1_z(u)$ in $u$ becomes $y^1_z$ for one value of $z\in\{z_1,\dots,z_L\}$). The next theorem summarizes the results of Sections \ref{sy} to \ref{pi}.
\begin{Theorem} Under Assumptions (M), (C1), (C2) and (G) (or (G') when $Z$ and $W$ are binary), $u_Y$ is identified and so is $(\varphi^1_{z_\ell}(u))_{\ell=1}^L$ for all $u\in [0,u_Y)$.
\end{Theorem}

\subsection{Partial identification}
\label{sec.paid}

When $u> u_Y$, it is possible to partially identify $(\varphi^1_{z_\ell}(u))_{\ell=1}^L$. Indeed, \eqref{change} becomes $$\sum_{\ell=1}^L \P(\varphi_{z_\ell}^1(U)\ge \varphi^1_{z_\ell}(u),Z=z_\ell|W=w_k)\ge \sum_{\ell=1}^L \P(U\ge u,Z=z_\ell|W=w_k)$$ because $\{U\ge u\}\subset\{\varphi_{z}^1(U)\ge \varphi^1_{z}(u)\}$ by (weak) monotonicity of $\varphi_z^1$. As a result, we have 
$\sum_{\ell=1}^LS^1(\varphi_{z_\ell}^1(u),z_\ell|w_k)\ge  u\, \text{for } k=1,\dots,K,$
but because $S^1(\cdot,z|w)$ is only identified on $[0,c_z]$ we cannot directly use these equations. As $S^1(t\wedge c_{z,w},z|w)\ge S^1(t,z|w)$ for all $t\in \R$, we leverage instead the following proposition which gives an outer set to the identified set.

\begin{Proposition}\label{PI}
Under Assumption (M), for all $u> u_Y$, we have 
\begin{equation}
\label{outer set}
(\varphi^1(z_\ell,u))_{\ell=1}^L\in\left\{\theta\in [0,\infty]^L\Big|\theta \notin\prod_{\ell=1}^L[0,y^1_{z_\ell}),\min_{k=1}^KR_{k,u}(\theta)\ge 0\right\},
\end{equation}
where $R_{k,u}(\theta)=\sum_{\ell=1}^LS^1(\theta_\ell\wedge c_{z_\ell}, z_\ell|w_k)-(1-u)$.
\end{Proposition}

\begin{figure}[ht]
\centering
\begin{tikzpicture}
\fill[color=myblue]
 (0,1.5) -- (0,2)
-- (0,2) -- (2,2)
-- (2,2) -- (2,0)
-- (2,0) -- (1,0)
-- (1,0) -- (1,1.5)
-- (1,1.5) -- (0,1.5) -- cycle ;
\draw[red] (1, 2) .. controls (1.15,1.55) .. (2,0.5);
\draw[->] (0,0) -- (2,0);
\draw (1,0) node[below] {$y^1_0$};
\draw (1,-0.1) -- (1,0.1);
\draw [->] (0,0) -- (0,2);
\draw (0,1.5) node[left] {$y^1_1$};
\draw (-0.1,1.5) -- (0.1,1.5);
\draw (1,-1) node{(i)};
\end{tikzpicture}
\begin{tikzpicture}
\fill[color=myblue]
 (0,1.5) -- (0,2)
-- (0,2) -- (0.5,2)
-- (0.5,2) -- (0.5,1.5)
-- cycle ;
\fill[color=myblue]
 (1,0) -- (2,0)
-- (2,0) -- (2,0.5)
-- (1,0.5) --(1,0)
-- cycle ;
\draw[red] (0.35, 2) .. controls (0.65,0.5) .. (2,0.35);
\draw[->] (0,0) -- (2,0);
\draw (1,0) node[below] {$y^1_0$};
\draw (1,-0.1) -- (1,0.1);
\draw [->] (0,0) -- (0,2);
\draw (0,1.5) node[left] {$y^1_1$};
\draw (-0.1,1.5) -- (0.1,1.5);

\draw (1,-1) node{(ii)};
\end{tikzpicture}
\begin{tikzpicture}

\fill[color=myblue]
 (1,0) -- (2,0)
-- (2,0) -- (2,0.5)
-- (1,0.5) --(1,0)
-- cycle ;
\draw[red] (0, 0.85) .. controls (0.60,0.60) .. (2,0.25);
\draw[->] (0,0) -- (2,0);
\draw (1,0) node[below] {$y^1_0$};
\draw (1,-0.1) -- (1,0.1);
\draw [->] (0,0) -- (0,2);
\draw (0,1.5) node[left] {$y^1_1$};
\draw (-0.1,1.5) -- (0.1,1.5);
\draw (1,-1) node{(iii)};
\end{tikzpicture}
\begin{tikzpicture}
\fill[color=myblue]
 (0,1.5) -- (0,2)
-- (0,2) -- (0.5,2)
-- (0.5,2) -- (0.5,1.5)
-- cycle ;
\draw[red] (0.4, 2) .. controls (0.7,0.7) .. (0.97,0);
\draw[->] (0,0) -- (2,0);
\draw (1,0) node[below] {$y^1_0$};
\draw (1,-0.1) -- (1,0.1);
\draw [->] (0,0) -- (0,2);
\draw (0,1.5) node[left] {$y^1_1$};
\draw (-0.1,1.5) -- (0.1,1.5);
\draw (1,-1) node{(iv)};
\end{tikzpicture}
\captionsetup{labelsep=none}
\caption{. Outer set (in grey) in the four different cases and the set $\{\theta \in [0,\infty]^L |  \min\limits_{k=1}^K \sum_{\ell=1}^LS^1(\theta_\ell, z_\ell|w_k)= 1-u \}$ (in red).}
\label{fig:red}
\end{figure}
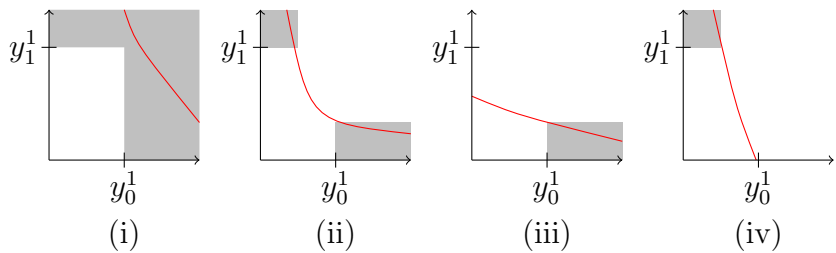

This outer set is not sharp because it does not take into account the constraints of continuity and monotonicity of $\varphi^1$. Let us now discuss how to compute this outer set as a finite union of product of intervals. We begin with the case where $Z$ is binary with support $\{0,1\}$. The set can have four types of shapes given in Figure~\ref{fig:red}:
\begin{itemize}
\item[] \begin{tabular}{ll}
(i) & the whole positive quadrant $[0,\infty]^2$ with the exception of $[0,y^1_0)\times[0,y^1_1)$ \\
(ii) & $[0,\bar\theta_1] \times [y^1_1,\infty] \cup [y_0^1,\infty] \times [0,\bar\theta_2]$ \\
(iii) & $[y_0^1,\infty] \times [0,\bar\theta_2]$ \\
(iv) & $[0,\bar\theta_1] \times [y^1_1,\infty]$ \\[.2cm]
\end{tabular}
\end{itemize}
for some $\bar\theta_1,\bar \theta_2\ge 0$.

Now, we clarify how to obtain Figure~\ref{fig:red}. Take $\theta \in[0,\infty]^2$ outside $[0,y^1_0)\times[0,y^1_1)$. Remark that $S^1(t\wedge c_{z},z|w)$ does not depend on $t$ on $[y^1_{z},\infty)$. Indeed, if $y^1_{z}=t^1_{z}$ (i.e. $t^1_{z}\le c_{z}$), then $t\ge t^1_{z}$ and therefore $S^1(t\wedge c_{z,w},z|w)=S^1(t^1_{z},z|w)$ because $t^1_z$ is larger than the upper bound of the distribution of $T$ given $\{E=1,Z=z,W=w\}$. If instead $y^1_{z}=c_{z}$ (i.e. $c_z\le t^1_z$), we have $S^1(t\wedge c_{z},z|w)=S^1(t^1_{z},z|w)$. As a result, $\theta$ is in the outer set if and only if $(\theta_1+ (y^1_0-\theta_1)I(\theta_1>y^1_0), \theta_2+ (y^1_1-\theta_2)I(\theta_2>y^1_1))^\top$ belongs to the outer set. Therefore, it is enough to study the value of $\theta \mapsto \min_{k=1}^K R_{k,u}(\theta)$ on $[0,y^1_0] \times \{y^1_1\}\cup\{y^1_0\}\times [0,y_1^1]$ to draw the outer set.  We begin with $[0,y^1_0)\times \{y^1_1\}$. Because $S^1(\cdot ,0|w)$ is continuously decreasing on $[0,y^1_0)$, the set of vectors $(\theta_1,y^1_1)^\top$
in $[0,y^1_0)\times \{y^1_1\}$ such that $\min_{k=1}^K R_{k,u}(\theta)\ge 0$ is either a segment $[0,\bar{\theta}_1]\times\{y^1_1\}$  where $\bar{\theta}_1\in[0,y^1_1]$ or empty. If this set is not empty, as $S^1(\cdot\wedge c_{0} ,0|w)$ is constant on $[y^1_0,\infty]$, $[0,\bar{\theta}_1]\times[y^1_1,\infty]$ is in the outer set. If rather this set is empty, $[0,y^1_1)\times[y^1_1,\infty]$ is not part of the outer set. Then, implement the approach on $ \{y^1_0\}\times [0,y^1_1)$. Finally, when $\min_{k=1}^K R_{k,u}((y_0^1,y^1_1)^\top)\ge 0$, $[y_0^1,\infty]\times [y^1_1,\infty]$ is in the outer set too.

If $L\ge 3$, one should use a recursive procedure. The outer set in dimension $L$ can be computed as the union of the extrapolations of $L$ outer sets in dimension $L-1$. Indeed, one can begin with fixing the first coordinate $\theta_1$ to $y^1_{z_1}$ and compute the outer set for the other $L-1$ coordinates.  If this set is not empty, it should be extrapolated by allowing $\theta_1$ to belong to $[y^1_{z_1},\infty]$. In the same manner, one should use this approach for the $L-1$ other coordinates. The outer set is then the union of the $L$ sets obtained by fixing each coordinate.

\section{Estimation}
\label{sec.est}

\subsection{Estimation procedure}

We consider estimation with an i.i.d.\ sample of size $n$, $\{Y_i, \delta_i E_i, Z_i,W_i\}_{i=1}^n$. The goal is to estimate $ (\varphi^1_{z_\ell}(u))_{\ell=1}^L$ for $u< u_Y$, that is at quantiles where the regression function is point identified. 

We assume that we have consistent estimators $\widehat{S}^1$ of $S^1$, $\widehat{u}_Y$ of $u_Y$ and $\widehat{y}^1_z$ of $y^1_z$ for $z\in \{z_1,\dots,z_L\}$. Choices of these estimators are discussed in Sections \ref{sec.choice} and \ref{estbaru}. 

We introduce further notations. For $u\in [0,u_Y)$, let $V(u)$ be a positive definite $K\times K$ weighting matrix. For a $K\times K$ matrix $\bar V$ and a vector $v\in \mathbb{R}^K$, we define $\|v\|_{\bar V}^2=\sqrt{v^{\top}\bar Vv}$.  The estimator of $ (\varphi^1_{z_\ell}(u))_{\ell=1}^L$ is defined by
\begin{equation}\label{prog}  
(\widehat{\varphi}^1_{z_\ell}(u))_{\ell=1}^L \in \argmin_{\theta \in\prod_{\ell=1}^L[0,\widehat{y}_{z_\ell}^1)} \Big\|\Big( \sum_{k=1}^K\widehat{S}^1(\theta_\ell,z_\ell|w_k)-(1-u)\Big)_{k=1}^K\Big\|_{V(u)}^2,
\end{equation}
and only the values of $u$ such that $u< \widehat{u}_Y$ should be reported. In practice, we cannot compute the estimator at all $u\in[0,1]$. Instead, we choose a grid $0\le u_1<u_2,<\dots<u_M\le 1$, $M$ values at which we want to estimate $\varphi^1$. Take, for instance, $\{1/M,2/M,\dots,1\}$.

When there are no competing risks, the estimator defined in \eqref{prog} is the same as the one in \citet{beyhum2021nonparametric}, except that in the latter paper the term $1-u$ in \eqref{prog} is replaced with $e^{-u}$ because a different normalization of the distribution of $U$ is used. When there are competing risks, the main difference between the estimator of the present paper and the one of \citet{beyhum2021nonparametric} is the fact that here the function $S^1$ is the survival function of a subdistribution while in \citet{beyhum2021nonparametric} it is the survival function of the duration of interest. In the present paper this function can be estimated by smoothing the Aalen-Johansen estimator (see Section \ref{sec.choice} below), while in \citet{beyhum2021nonparametric} it is estimated by smoothing the Kaplan-Meier (Beran) estimator. This does not change the theoretical properties of solutions to \eqref{prog}, because the Kaplan-Meier estimator of the survival function and the Aalen-Johansen estimator of the survival of the subdistribution of dying from cause $1$ have similar asymptotic properties. Therefore, in the present paper, we skip the theoretical analysis of solutions to \eqref{prog} and refer the reader to \citet{beyhum2021nonparametric} for properties, proofs and further discussion. One major remaining difference between the estimation procedure in \citet{beyhum2021nonparametric} and the one in the present paper is that, here, we provide a theoretical analysis of how to estimate $u_Y$, see Section \ref{estbaru} below.

\subsection{Estimation of $S^1$}
\label{sec.choice}

In this subsection, we discuss choices of $\widehat{S}^1$ that work under competing risks and random right censoring. Remark that $S^1(t,z\vert w)=S^1(t\vert z,w)p_{z,w}$ where $p_{z,w}=\P(Z=z\vert W=w )$.  We define the following stochastic processes: 
\begin{align*}
& N_{z,w}^1(t)= \sum_{i=1}^nI(Y_i\le t, \delta_iE_i=1, Z_i=z,W_i=w) \\
& N_{z,w}(t) =\sum_{i=1}^nI(Y_i\le t, \delta_iE_i\ne 0, Z_i=z,W_i=w) \\
& Y_{z,w}(t)=\sum_{i=1}^nI(Y_i\ge t,Z_i=z,W_i=w) \\
& Y_{z,w}= \sum_{i=1}^nI(Z_i=z,W_i=w) \quad \mbox{and} \quad Y_{w}=\sum_{i=1}^nI(W_i=w).
\end{align*}
We estimate $S^1(\cdot\vert z,w)$ using the Aalen-Johansen estimator of the cause-specific survival function (see \citet{aalen1978empirical,geskus2020competing}), that is
$$
\widehat{S}_{AJ}^1(t\vert z, w)=1-\sum_{s\le t} \left(\prod_{u\le s}\left[1-\frac{dN_{z,w}(u)}{Y_{z,w}(u)}\right]\right) \frac{dN^1_{z,w}(s)}{Y_{z,w}(s)}.$$
To ensure that \eqref{prog} has a unique solution, one should smooth $\widehat{S}^1_{AJ}$. Various techniques are available in the literature, including local polynomials and kernel smoothing. For instance, concerning the latter, if we use a kernel $K$ with a bandwidth $\epsilon$, we obtain
$$\widetilde{S}^1(t\vert z,w)=\int \widehat{S}^1_{AJ}(t-s\epsilon\vert z, w)K(s)ds.$$
Our final estimator of $S^1$ is $$\widehat{S}^1(t,z\vert w)= \widetilde{S}^1(t\vert z,w)\widehat{p}_{z,w},$$ where $\widehat{p}_{z,w} =Y_{z,w}/Y_w$.

\subsection{Estimation of $\boldsymbol{y}^1_z$ and $\boldsymbol{u_Y}$}
\label{estbaru}

Let us now introduce estimators of $y^1_z,z=z_1,\dots,z_L$ and $u_Y$. These estimators are new and not discussed in \citet{beyhum2021nonparametric}, which is why we provide theoretical results. Because $y^1_z$ is the upper bound of the support the distibution of $Y$ given $\{E=1,Z=z\}$, a natural estimator is 
$$\widehat{y}^1_z=\max\limits_{i\in\{1,\dots,n\}:\ Z_i=z,\ \delta_iE_i=1}Y_i.$$
We have the following result.

\begin{Proposition}\label{constauell}
$\widehat{y}^1_z\xrightarrow{\P}y^1_z$ for all $z\in\{z_1,\dots,z_L\}$.
\end{Proposition}

The proof is given in Appendix B. In turn, the proposed estimator of $u_Y$ is $\widehat{u}_Y=u_{\widehat{m}_Y}$, where
\begin{equation}\label{estimator_uy} 
\widehat{m}_Y =\argmin \Big\{m\in\{1,\dots, M\}:\ \exists\ell\in\{1,\dots,L\}\text{ such that }\widehat{\varphi}^1_{z_\ell}(u_m)\ge \widehat{y}^1_{z_\ell}-\Delta_\ell\Big\},
\end{equation}
for some small $\Delta_\ell>0,\ell=1,\dots, L$. The rationale of this estimator is as follows. By definition, $u_Y$ is the lowest value of $u$ such that there exists $\ell  \in \{1,\dots,L\}$ for which the left limit of $\varphi^1_{z_\ell}(\cdot)$ in this value is equal to $y^1_{z_\ell}$. If $\widehat{y}^1_z$ is consistent, then $\widehat{y}^1_z-\Delta$ should be close to $y^1_z$. As a result, since $\widehat{\varphi}^1$ is consistent, $\widehat{u}_Y$ is close to $u_Y$. The main role of $\{\Delta_\ell\}_{\ell=1}^L$ is to provide a cushion which ensures that the probability of $\{\widehat{u}_Y>u_Y\}$ is small. This event is not desirable because it would lead the researcher to report results for values of $u$ for which $(\varphi^1_{z_\ell}(u))_{\ell=1}^L$ is not identified. The following proposition formalizes these ideas.

\begin{Proposition}\label{consbaru}
Assume that $\widehat{\varphi}^1_{z}(\cdot)$ is uniformly consistent on $[0,u_Y)$ with rate of convergence $r_n\to 0$, that is 
$$ \sup\limits_{z=z_1,\dots,z_L,\ u\in[0,u_Y)}|\widehat{\varphi}^1_z(u)-\varphi^1_z(u)|=O_P(r_n)$$ and that the grid (which depends on $n$) is dense in the sense that
$$\lim_{n\to\infty}\max_{u\in[0,1]} \, \min_{u_1,\ldots,u_M} |u-u_m|\to 0.$$
We also choose the sequences $\Delta_\ell=(\Delta_\ell)_n,\ \ell =1,\dots, L$ such that $r_n/\min_{\ell=1}^L\Delta_\ell\to 0$.
Then, we have $\widehat{u}_Y\xrightarrow{\P}u_Y$ and $\lim\limits_{n\to\infty}\P(u_{\widehat{m}_Y-1}\le u_Y)\to 1$ (where $u_{\widehat{m}_Y-1}=0$ if $\widehat{m}_Y=1$).
\end{Proposition}

The proof is given in Appendix B. An important remark is that the theoretical analysis in \citet{beyhum2021nonparametric} concerns the ideal estimator 
\begin{equation}\label{progideal}  
(\widetilde{\varphi}^1_{z_\ell}(u))_{\ell=1}^L \in \argmin_{\theta \in\prod_{\ell=1}^L[0,y^1_{z_\ell})} \Big\|\Big( \sum_{k=1}^K\widehat{S}^1(\theta_\ell,z_\ell|w_k)-(1-u)\Big)_{k=1}^K\Big\|_{V(u)}^2
\end{equation}
for all $u< u_Y$ where it is assumed that $y^1_z,z=z_1,\dots,z_L$ and $u_Y$ are known. Our estimator $\widehat{\varphi}^1$ and $\widetilde{\varphi}^1$ coincide on the event 
$$\mathcal{E}=\left\{u \in [0,1] : (\widetilde{\varphi}^1_{z_\ell}(u))_{\ell=1}^L\in \prod_{\ell=1}^L[0,\widehat{y}_{z_\ell}^1), u< \widehat{u}_Y\wedge u_Y\right\}.$$
For $u<u_Y$, when $(\widetilde{\varphi}^1_{z_\ell}(u))_{\ell=1}^L$ is consistent, we have $\lim\limits_{n\to \infty}\P(\mathcal{E})=1$ because $\widehat{y}_{z}^1, z=z_1,\dots,z_L$ and $\widehat{u}_Y$ are consistent. Remark that $\widehat{\varphi}^1=\widetilde{\varphi}^1I(E)+\widehat{\varphi}^1 (1-I(E))$ and $(\widehat{\varphi}^1_{z_\ell})^L_{\ell=1}\in\prod_{\ell=1}^L[0,\widehat{y}_{z_\ell}^1)$ is bounded because $\widehat{y}_{z}^1\le y^1_z, z=z_1,\dots,z_L$ almost surely. Hence, by the continuous mapping theorem and Slutsky's theorem, $\widehat{\varphi}^1$ has the same asymptotic properties as $\widetilde{\varphi}^1$ and the results of \citet{beyhum2021nonparametric} are not altered.

\section{Simulations}
\label{sec.sim} 

Let us consider the following data generating processes (henceforth, DGPs). The variable $U$ has a uniform distribution on the interval $[0,1]$ and $W$ is a Bernoulli random variable with parameter $2/3$, independent of $U$.  We generate $$Z=I\left(4U+\epsilon-1\ge 0\right)W,$$
 where $\epsilon\sim\mathcal{N}(0,1)$ is independent of $(U,W)$. Hence, in this simulation experiment there is one-sided noncompliance as in our empirical application (see Section \ref{sec.app}). We set $E=I(U>p_Z)+1$, where $p_0=1/2$, $p_1=3/4$. The duration is 
$$ T=\left\{\begin{array}{cl} \varphi_1(Z,U)&\text{ if } U<p_Z\\
 \varphi_2(Z,U) &\text{ if } U>p_Z\end{array}\right.;\ $$
$$ \varphi_1(z,u)=\left\{\begin{array}{cl} 2u&\text{ if } z=0\\
u &\text{ if } z=1\end{array}\right.;\ \quad \varphi_2(z,u)=\left\{\begin{array}{cl} u-p_0&\text{ if } z=0\\
2(u-p_1) &\text{ if } z=1.\end{array}\right. $$
This leads to 
$$\varphi^1_0(u)=\left\{\begin{array}{cl} 2u&\text{ if } u\le \frac12\\
\infty &\text{ otherwise }\end{array}\right.;\  \quad \varphi^1_1(u)=\left\{\begin{array}{cl} u&\text{ if } u\le \frac34\\
\infty &\text{ otherwise }.\end{array}\right.$$
The treatment increases the probability that $E=1$ and reduces the duration until cause $1$ happens for a given value of $u$. Note that the support of $U|Z$ is $[0,1]$ because of the presence of the random noise $\epsilon$. Therefore, we have $t^1_0= 1$, $t_1^1=3/4$ and $u_E=1/2$. We propose two designs for the censoring variable:
\begin{align*}
\text{Design 1:}& \text{ $C$ is independent of $(U,Z,W)$ and uniform on the interval $[1/3,2/3]$;}\\
\text{Design 2:}& \text{ $C$ is independent of $(U,Z,W)$ and uniform on the interval $[1/3,3/2]$.}
\end{align*}
In the first experiment, we have $u_C= 1/3<u_E$, hence identification fails for $u>u_Y=1/3$ because of censoring. In the second exercise, it holds that $u_C=1$ and partial identification arises at $u_Y=1/2$ due to competing risks.

With these DGPs, the probability of treatment given $W=1$ is $\P(Z=1|W=1)=73\%$. Under design 1, $30\%$ of the observations are censored, while under design 2, it is only $10\%$ (all these quantities are averages over 1,000,000 Monte Carlo replications). The sample size was set at $n=10,000$. We generated $1,000$ replications of the model under both designs. The estimator \eqref{prog} was computed on the grid $u_m=0.01m,\ m=1,\dots,100.$ We used a local polynomial of degree $1$ with an Epanechnikov Kernel to smooth the Aalen-Johansen estimator of the cause-specific survival function. The bandwidth was selected according to the usual rule of thumb for normal densities. To estimate $u_Y$, we used the estimator defined in \eqref{estimator_uy}. The quantity $\Delta_\ell$ was chosen as the optimal bandwidth for normal density estimation with Epanechnikov kernel of the sample $\{Y_i:\ i\in\{1,\dots,n\} \text { such that } Z_i=z_\ell,\ \delta_iE_i=1\}$.  

\begin{figure}[H]
\begin{minipage}{0.48\textwidth}
  \centering
  \includegraphics[width=80mm]{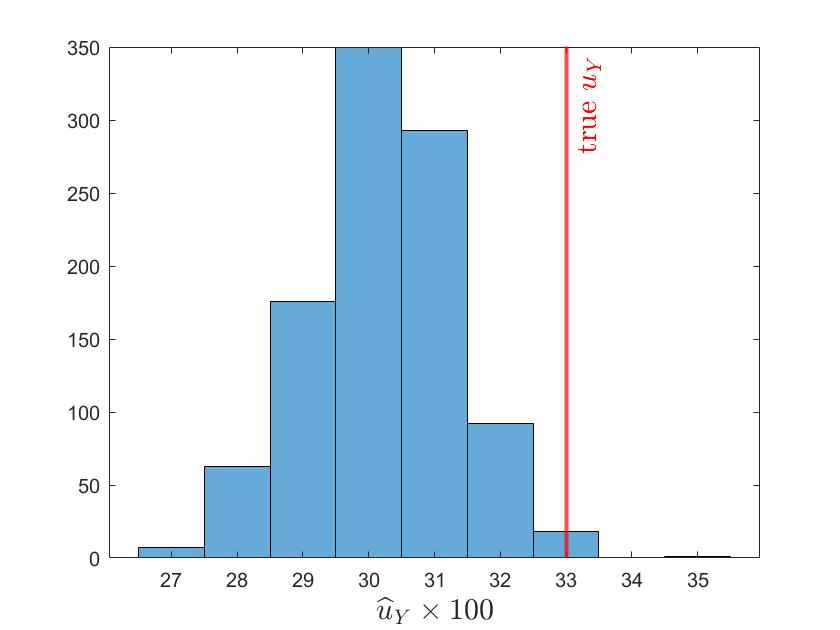}
 
  \caption{Values of $\widehat{u}_Y$ for 1,000 replications of the DGP under design 1.}
   \label{fig:hist_design1} 
    \end{minipage}
    \quad\quad
    \begin{minipage}{0.48\textwidth}
        \centering
   \includegraphics[width=80mm]{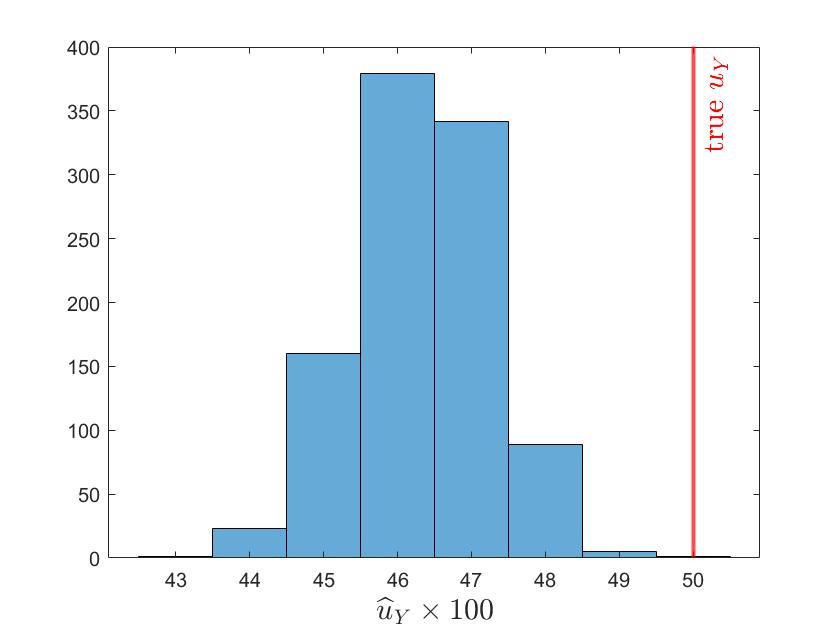}
  \caption{Values of $\widehat{u}_Y$ for 1,000 replications of the DGP under design 2.}
      \label{fig:hist_design2}
      \end{minipage}
\end{figure}

\begin{figure}[H]
\begin{minipage}{0.48\textwidth}
  \centering
  \includegraphics[width=80mm]{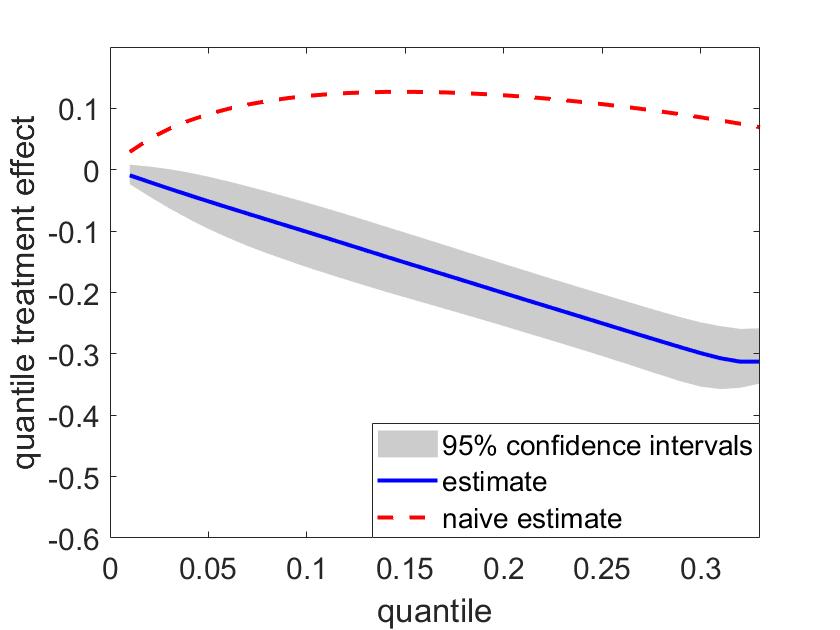}
 
  \caption{Average estimated quantile treamment effect under design 1.}
   \label{fig:CI_design1} 
    \end{minipage}
    \quad\quad
    \begin{minipage}{0.48\textwidth}
        \centering
   \includegraphics[width=80mm]{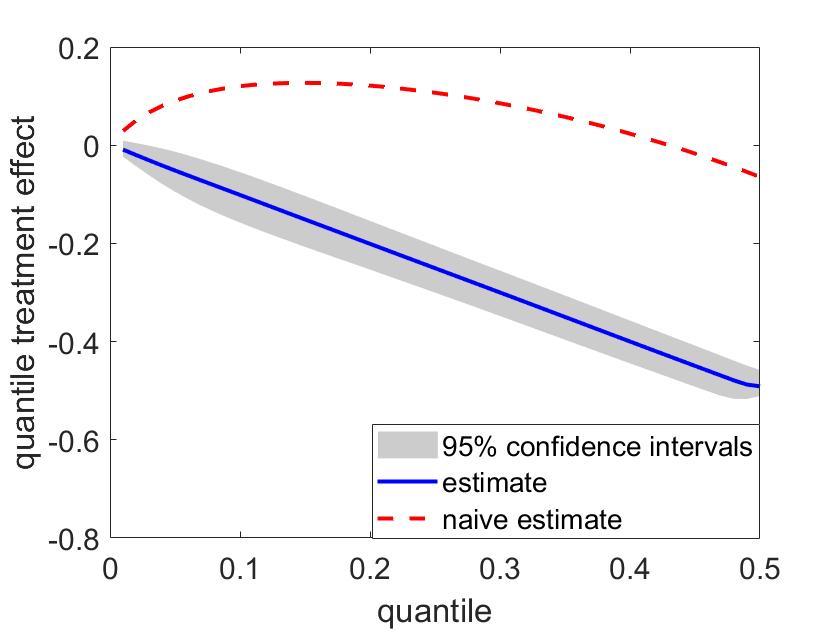}
  \caption{Average estimated quantile treatment effect under design 2.}
      \label{fig:CI_design2}
      \end{minipage}
\end{figure}

\begin{figure}[H]
\begin{minipage}{0.48\textwidth}
  \centering
  \includegraphics[width=80mm]{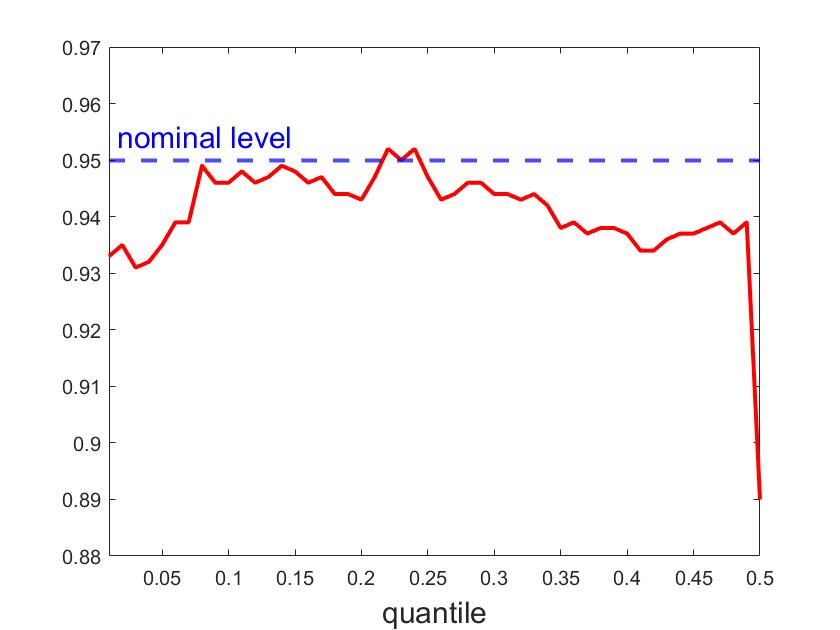}
 
  \caption{Coverage of 95\% confidence intervals  under design 1.}
   \label{fig:cov_design1} 
    \end{minipage}
    \quad\quad
    \begin{minipage}{0.48\textwidth}
        \centering
   \includegraphics[width=80mm]{coverage_design2.jpg}
  \caption{Coverage of 95\% confidence intervals under design 2.}
      \label{fig:cov_design2}
      \end{minipage}
\end{figure}

Figures ~\ref{fig:hist_design1} and ~\ref{fig:hist_design2} display the histograms of the values of $\widehat{u}_Y$ for design 1 (left) and 2 (right). The estimator $\widehat{u}_Y$ almost always yields a quantile lower than but close to $u_Y$, which avoids reporting results at points at which $\varphi^1$ is not identified. In Figures ~\ref{fig:CI_design1} and ~\ref{fig:CI_design2}, we present the average value of our estimator of the quantile treatment effect $\widehat{\varphi}^1_1-\widehat{\varphi}^1_0$ for points of the grid $\{u_m\}_{m=1}^M$ smaller than $u_Y$. We also show the average of the naive estimator of the quantile treatment effect, which estimates $\varphi^1_z$ by inverting the Aalen-Johansen estimator of the survival function $\P(T\ge t, E=1|Z=z)$. The naive estimator directly compares treated observations to untreated ones. It ignores the endogeneity issue and, hence, is biased. The figures also exhibit the average of the bounds of the $95\%$ confidence intervals of the quantile treatment effects computed with 200 bootstrap draws.  The true quantile treatment effects are not reported because they are indistinguishable from our estimator. Finally, Figures ~\ref{fig:cov_design1} and ~\ref{fig:cov_design2} show that the coverage of these 95\% confidence intervals is almost nominal.

\section{Real data application}
\label{sec.app}

We apply our methodology to the Health Insurance Plan of Greater New York experiment. This clinical trial aimed to evaluate the effect of periodic screening examinations (which aim to detect breast cancer) on breast cancer mortality. The study started in 1963 and lasted until 1986. The experiment follows 60,695 women between 40 and 60 years old. About half (30,565) of the participants were randomized into the control group ($W=0$) while the rest (30,100) were assigned to the intervention group ($W=1$). Members of the intervention arm were offered a treatment consisting of an initial breast examination and mammography and three yearly subsequent screens.  9,984 participants randomized into the intervention group refused the treatment, which corresponds to a rate of non-compliance of 33\%. The women in the study group that accepted the treatment exhibited largely different observable characteristics from the ones that refused screening (see \citet{shapiro1997periodic}). This suggests that the treatment is endogenous. Let $Z$ be the variable which equals $1$ for women who were offered and accepted the treatment and $0$ otherwise. 

All participants were followed through three mail surveys, respectively, 5, 10 and 15 years after their entry into the study. The outcome duration of interest $T$ is the time from the initial randomization into the study until death. Censoring arises because some subjects are lost to follow-up before they die. Hence, $C$ is the time between registration into the experiment and the last response to a follow-up survey.  We consider two competing risks: deaths from breast cancer ($j=1$) and death from any other cause ($j=2$). In the sample, there are $786$ deaths from breast cancer, $13,798$ deaths from other causes and $46,111$ censored observations.

\begin{figure}[H]
\begin{minipage}{0.48\textwidth}
  \centering
  \includegraphics[width=80mm]{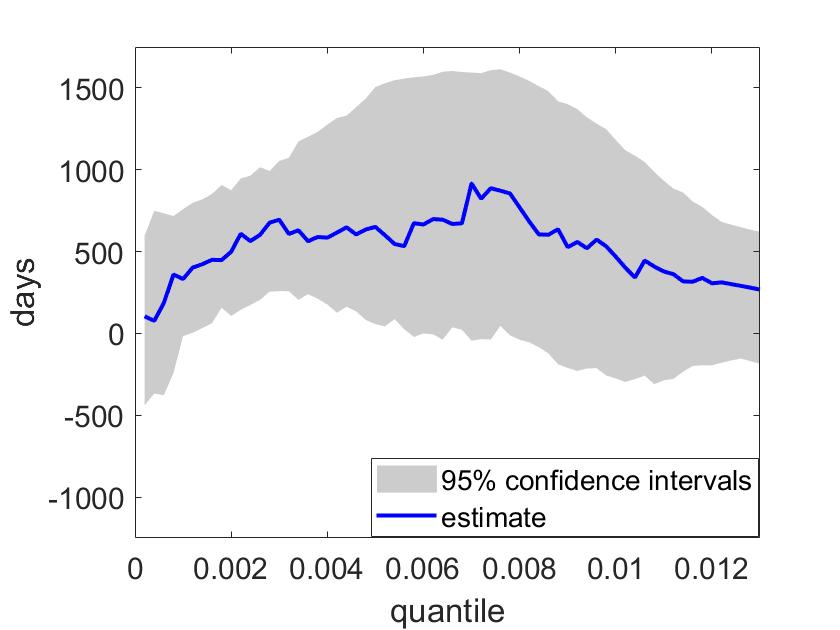}
  \caption{Quantile treatment effects on the subdistribution of the duration until death caused by breast cancer.}
   \label{fig:QTE_bc} 
    \end{minipage}
    \quad\quad
    \begin{minipage}{0.48\textwidth}
        \centering
   \includegraphics[width=80mm]{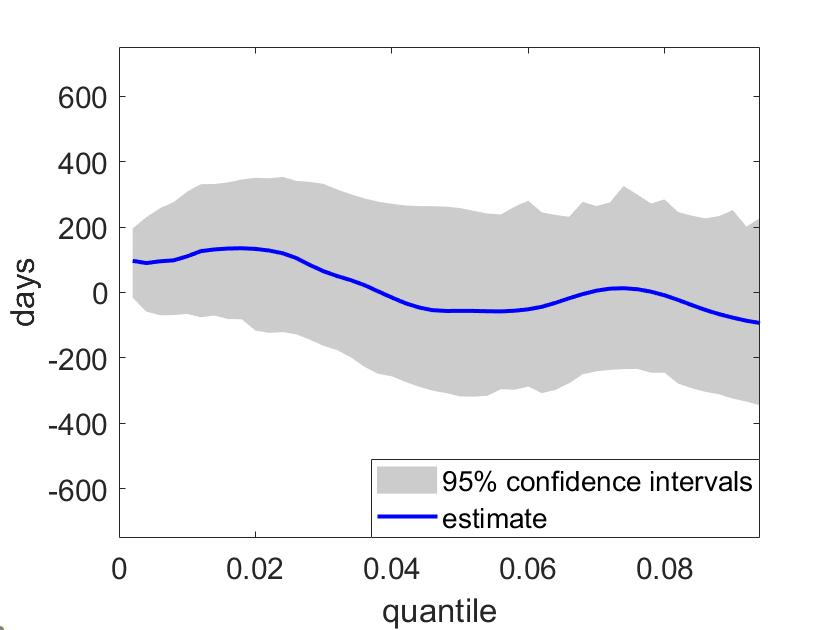}
  \caption{Quantile treatment effects on the subdistribution of the duration until death by another cause than breast cancer.}
      \label{fig:QTE_o}
      \end{minipage}
\end{figure}

As in the simulations, the Aalen-Johansen estimators of the survival functions are smoothed using a local polynomial of degree $1$. The bandwidths are selected similarly. The confidence intervals were computed using 200 bootstrap replications. For death by breast cancer, we computed the results on a grid of values of $U$ starting at $u_1=2\times 10^{-4}$ with step $2\times 10^{-4}$. The value of the estimator of $u_Y$ was $0.013$, corresponding to the 1.3\% quantile. The quantile treatment effects can only be estimated for such low quantiles because the value of $T^1$ is almost always infinity (most women do not die from breast cancer) and censored (most women do not die in the 15 years following screening). The estimates of the quantile treatment effects along with the 95\% confidence intervals are reported in Figure ~\ref{fig:QTE_bc}. The quantile treatment effects are insignificant except for some low quantiles. This seems in contrast with findings in \citet{shapiro1997periodic}, who concluded that the treatment reduced the probability of dying from breast cancer before a certain time, but in the latter paper the endogeneity issue is ignored. For cause $2$ (death by other causes than breast cancer), a grid of values of $U$ starting at $u_1=0.02$ with step $0.02$ was chosen. We found $\widehat{u}_Y=0.12$. However, the estimated quantile treatment effects close to 
$\widehat{u}_Y$ exhibit a very irregular behavior. Hence, we chose to present in Figure ~\ref{fig:QTE_o} only the results for quantiles between $0$ and $9.8\%$. As expected, the treatment does not have a significant effect on the subdistribution of the time until of death from another cause than breast cancer.

\section{Conclusion}
This paper exhibits the link between competing risks models and nonparametric regression models in the presence of endogeneity. Thanks to this relationship, we are able to formulate our problem in terms of a quantile instrumental regression model. We study identification and estimation of the model. Numerical experiments assess the small sample performance of the method. We show how to revisit an empirical application using this paper's approach.

Many possible directions for future research are interesting. A valuable generalization would allow for continuous or even dynamic treatments, instrumental variables and covariates $X$. With continuous variables, this could be done by replacing the system of equations (\ref{Sy}) by a (potentially infinite) set of integral equations
$$\int S^1(\varphi^1_z(u),z,x|w)dz=1-u, \text{ for all $x,w,u$ in the support of $X$, $W$ and $U$}.$$
This is an ill-posed problem and regularization of the estimator would be required. 

It would also be interesting to identify directly the properties of the marginal distributions of the risks. Following this goal, one could assume that the risks are independent conditional on a set of covariables. Under this condition, the competing events can be treated as independent censoring. Another approach could be to extend models from the dependent censoring literature (such as in \citet{basu1978identifiability,emoto1990weibull, deresa2019semiparametric,czado2021}) to the case where endogeneity is allowed.

Finally, it should be noted that because the duration $T^1$ has a mass at infinity, all the discussion of this paper can easily be adapted to the estimation of the latency in a cure model with endogenous treatment. In fact a competing risks model can be seen as a cure model where the cure status is known for some observations (see \citet{betensky2001nonparametric}). A promising project could investigate the identification of the cure fraction (the cause-specific probability of failure in the context of competing risks) when the treatment is endogenous. 

\label{sec.conc}
\bibliographystyle{dcu}
\bibliography{paper-210427}
\appendix

\renewcommand{\theequation} {A.\arabic{equation}}

\section{Appendix A: On the model}
\label{sec.A}

\subsection{Distribution of $U$} We prove the following Lemma.
\begin{Lemma}\label{A1}
$U$ has a uniform distribution on the interval $[0,1]$.
\end{Lemma}
\begin{Proof}
For $u\in[0,1]$, we have
\begin{align}
\notag\P( U\le u)&=\P\Big(F_{1}(U_1)p\le u\Big|E=1\Big)p+\P\Big(p+(1-p)F_{2}(U_2)\le u\Big|E=2\Big)(1-p)\\
\notag &= \P\Big(F_{1}(U_1)\le \frac{ u}{p}\Big|E=1\Big)p+\P\Big(F_{2}(U_2)\le \frac{ u-p}{1-p}\Big|E=2\Big)(1-p)\\
\label{lastdistrib}&=p\Big(\frac{ u}{p}I(u\le p) +I(u> p)\Big) +(1-p)\frac{ u-p}{1-p}I( u>p)= u,
\end{align}
where \eqref{lastdistrib} is because $F_{j}(U_j),j=1,2$ have a uniform distribution on the interval $[0,1]$.
\end{Proof}

\subsection{On the rank invariance assumption}
For a given event type $j\in\{1,2\}$, the structural model of Section \ref{sec:model} implies the following rank invariance property:
\begin{itemize}
\item[(RI)] For two subjects $i_1$ and $i_2$ and $z\in\{z_1,\dots,z_L\}$, if $T^j_{i_1}(z)<T^j_{i_2}(z)$, then $(T^j_{i_1}(z'),T^j_{i_2}(z'))=(\infty,\infty) \text{ or }T^j_{i_1}(z')<T^j_{i_2}(z')$ for all  $z'\in\{z_1,\dots,z_L\}$.
\end{itemize}
Indeed, $T^1_{i_1}(z)<T^1_{i_2}(z)$ implies that $U_{i_1}<U_{i_2}$ because $\varphi^1_z$ is increasing. Then, if $U_{i_1}<p_{z'}$, we have $T^1_{i_1}(z')<T^1_{i_2}(z')$ and otherwise it holds that $(T_{i_1}(z'),T_{i_2}(z'))=(\infty,\infty)$. The proof for $j=2$ is similar.

Let us discuss the implications of this assumption in a simple example. Let $Z$ be a treatment for COVID-19 taking the values $0$ (untreated) and $1$ (treated). $T$ is the duration until one of the two following competing risks happens: recovery ($j=1$) and death ($j=2$). We consider two subjects $i_1$ and $i_2$. If $T^1_{i_1}(0)<T^1_{i_2}(0)$, then $i_1$ recovers and $i_2$ dies or recovers after $i_1$. Table~\ref{fig:cure} summarizes the different treated potential outcomes for $(T^1_{i_1}(1),T^1_{i_2}(1))$ allowed by the rank invariance assumption in this case. When $T^2_{i_1}(0)<T^2_{i_2}(0)$, then $i_1$ dies and $i_2$ recovers or dies after $i_1$ and the possible treated potential outcomes for $(T^2_{i_1}(1),T^2_{i_2}(1))$  are displayed in Table~\ref{fig:death}. It can be seen that the rank invariance assumption allows for many natural treatment effects.
\begin{table}[H]
\begin{minipage}{0.48\textwidth}
\centering
\begin{tabular}{|l|c|c|}
\hline \diagbox{$\boldsymbol{i_1}$}{$\boldsymbol{i_2}$}& \textbf{recovers} & \textbf{dies} \\
\hline \textbf{recovers} & \begin{tabular}{@{}l@{}}YES ($i_1$ recovers before $i_2$)\\ NO (otherwise)\end{tabular} & YES\\
\hline  \textbf{dies} & NO & YES\\
\hline
\end{tabular}
\caption{Possible combinations of treated outcomes when $i_1$ recovers and $i_2$ dies or recovers after $i_1$ without treatment.}
\label{fig:cure} 
\end{minipage}
\hfill%
\begin{minipage}{0.48\textwidth}
\centering
{\begin{tabular}{|l|c|c|}
\hline \diagbox{$\boldsymbol{i_1}$}{$\boldsymbol{i_2}$}& \textbf{recovers} & \textbf{dies} \\
\hline\textbf{recovers} & YES & NO\\
\hline \textbf{dies} & YES & \begin{tabular}{@{}l@{}}YES ($i_1$ dies before $i_2$)\\ NO (otherwise)\end{tabular} \\
\hline
\end{tabular}}
\caption{Possible combinations of treated outcomes when $i_1$ dies and $i_2$ recovers or dies after $i_1$ without treatment.}
\label{fig:death}
      \end{minipage}
\end{table}

\section{Appendix B: Proof of estimation results}
\label{sec.B}

\subsection{Proof of Proposition \ref{constauell}}
Let us consider two cases. If $y^1_z<\infty$, for $\epsilon >0$, we have $ |\widehat{y}^1_z-y^1_z|> \epsilon$ if and only if $Y_i< y^1_z-\epsilon$ for all $i\in\{1,\dots,n\}$ such that $Z_i=z,\ \delta_iE_i=1$. This has a probability at most $\P(Y\le y^1_z-\epsilon)^{Y_{z}^1}$, where 
$$Y_{z}^1=\sum_{i=1}^nI(Z_i=z,\ \delta_iE_i=1).$$
By the law of large numbers, $Y_{z}^1/n$ converges to $\P(Z=z,\ \delta E=1)>0$ in probability. Hence, $Y_{z}^1$ goes to $\infty$ in probability and 
$$\P( |\widehat{y}^1_z-y^1_z|> \epsilon)\le \P(Y \le y^1_z-\epsilon)^{Y_{z}^1}\to 0,$$
because $\P(Y \le y^1_z-\epsilon)<1$, by definition of $y^1_z$, which shows that $\widehat{y}^1_z\xrightarrow{\P}y^1_z$.

If instead  $y^1_z=\infty$, for $M>0$, $\widehat{y}^1_z<M$ if and only if $Y_i< M$ for all $i\in\{1,\dots,n\}$ such that $Z_i=z,\ \delta_iE_i=1$. This  event has probability at most $\P(Y \le M)^{Y_{z}^1}$. Because $\P(Y < M)<1$ by definition of $y^1_z$, we obtain $\P(\widehat{y}^1_z<M)\to0$ and, hence, that $\widehat{y}^1_z\xrightarrow{\P}\infty$. \hfill $\Box$

\subsection{Proof of Proposition \ref{consbaru}}
Let us define $\widehat{u}^I_Y =\min\{u\in[0,1]:\ \exists\ell\in\{1,\dots,L\}\text{ such that }\widehat{\varphi}^1_{z_\ell}(u)\ge \widehat{y}_{z_\ell}^1-\Delta_\ell\}$, which is similar to $\widehat{u}_Y$ except that the minimization is not over the grid. Notice that $|\widehat{u}^I_Y-\widehat{u}_Y|\le \max\limits_{u\in[0,1]}\min\limits_{u_1,,\dots,u_M}|u-u_m|\to 0$ and therefore $\widehat{u}^I_Y-\widehat{u}_Y=o_P(1)$.

We introduce
 $u^*(\Delta)=\min_{\ell=1}^L(\varphi_{z_\ell}^1)^{-1}\Big(y^1_{z_\ell}-\frac{\Delta_\ell}{2}\Big)$ and $\ell^*(\Delta)$ such that $\varphi^1_{z_{\ell^*(\Delta)}}(u^*(\Delta))=y^1_{z_{\ell^*(\Delta)}}-\frac{\Delta_{\ell^*(\Delta)}}{2}$. We have 
\begin{align}
\notag \widehat{\varphi}^1_{z_{\ell^*(\Delta)}}(u^*(\Delta))&\ge \varphi^1_{z_{\ell^*(\Delta)}}(u^*(\Delta))-\sup\limits_{z=z_1,\dots,z_L,\ u\in[0,u_Y)}|\widehat{\varphi}^1_z(u)-\varphi^1_z(u)|\\
\notag &= y^1_{z_{\ell^*(\Delta)}}-\frac{\Delta_{\ell^*(\Delta)}}{2}-\sup\limits_{z=z_1,\dots,z_L,\ u\in[0,u_Y)}|\widehat{\varphi}^1_z(u)-\varphi^1_z(u)|\\
\label{compl}&\ge \widehat{y}^1_{z_{\ell^*(\Delta)}}-\frac{\Delta_{\ell^*(\Delta)}}{2}-\sup\limits_{z=z_1,\dots,z_L,\ u\in[0,u_Y)}|\widehat{\varphi}^1_z(u)-\varphi^1_z(u)|,
\end{align}
because $\widehat{y}^1_{z_{\ell^*(\Delta)}}\le  y^1_{z_{\ell^*(\Delta)}}$.
It holds that $\sup\limits_{z=z_1,\dots,z_L,\ u\in[0,u_Y)}|\widehat{\varphi}^1_z(u)-\varphi^1_z(u)|=O_P(r_n)$ and $r_n/\min_{\ell=1}^L\Delta_\ell\to 0$, which yields 
$$\P\left(\sup\limits_{z=z_1,\dots,z_L,\ u\in[0,u_Y)}|\widehat{\varphi}^1_z(u)-\varphi^1_z(u)|\le \frac{\min_{\ell=1}^L\Delta_\ell}{2}\right)\to 1.$$
This and \eqref{compl} imply that
$$\P\left(u^*(\Delta)\ge\widehat{u}^I_Y\right)\ge \P\left(\widehat{\varphi}^1_{z_{\ell^*(\Delta)}}(u^*(\Delta))\ge \widehat{y}^1_{z_{\ell^*(\Delta)}}-\Delta_{\ell^*(\Delta)}\right)\to 1,$$ 
and since $u_Y\ge u^*(\Delta)$, we obtain $\P(u_Y\ge \widehat{u}_Y^I)\to 1$. Notice that $u_{\widehat{m}_Y-1}\le \widehat{u}^I_Y$, therefore $\P(u_{\widehat{m}_Y-1}\le u_Y)\to 1$.

Next, take $\epsilon>0$. We define 
$u^{**}(\epsilon)=\min_{\ell=1}^L(\varphi_{z_\ell}^1)^{-1}(y_{z_\ell}^1-\epsilon)$. For $\ell\in\{1,\dots,L\}$, we have
\begin{align*}
\widehat{\varphi}^1_{z_{\ell}}(u^{**}(\epsilon))&\le \varphi^1_{z_{\ell}}(u^{**}(\epsilon))+\sup\limits_{z=z_1,\dots,z_L,\ u\in[0,u_Y)}|\widehat{\varphi}^1_z(u)-\varphi^1_z(u)|\\
&\le y_{z_\ell}^1-\epsilon+o_P(1).
\end{align*}
Since $\Delta \to 0$ and $\widehat{y}_{z_\ell}^1\xrightarrow{\P}y^1_{z_\ell}$, this leads to 
$$\P\left(\widehat{\varphi}^1_{z_{\ell}}(u^{**}(\epsilon))\le \widehat{y}_{z_\ell}^1-\Delta,\ \forall \ell\in\{1,\dots,L\}\right)\to 1.$$
Hence, $\P(u^{**}(\epsilon)\le \widehat{u}^I_Y)\to 1$. Since $\lim\limits_{\epsilon\to 0}u^{**}(\epsilon)=u_Y$, this and $\P(u_Y\ge \widehat{u}_Y^I)\to 1$ imply that $\widehat{u}^I_Y\xrightarrow{\P}u_Y$. We conclude using the fact that $\widehat{u}_Y-\widehat{u}^I_Y=o_P(1)$, which implies $\widehat{u}_Y\xrightarrow{\P}u_Y$. \hfill $\Box$

\end{document}